\documentclass[letter, twocolumn]{jpsj3}
\usepackage{physics, ifthen}

\newboolean{letter}
\setboolean{letter}{false}
\setboolean{letter}{true}

\usepackage{amsmath, amssymb}
\ifthenelse{\boolean{letter}}{\usepackage{newtxtext, newtxmath}}{}
\usepackage{graphicx}
\usepackage{bm}
\usepackage{multirow}
\usepackage {subfigmat}
\providecommand{\U}[1]{\protect\rule{.1in}{.1in}}

\ifthenelse{\boolean{letter}}{\bibliographystyle{jpsj}}{}

\ifthenelse{\boolean{letter}}{\title{Electric Multipoles of Double Majorana Kramers Pairs}}{}

\ifthenelse{\boolean{letter}}{\author{Yuki Yamazaki$^1$, Shingo Kobayashi$^2$, and Ai Yamakage$^1$}
	\inst{${}^1$Department of Physics, Nagoya University, Nagoya 464-8602, Japan
		\\
		${}^2$RIKEN Center for Emergent Matter Science, Wako, Saitama 351-0198, Japan
	}
}{}

\date{\today}

\ifthenelse{\boolean{letter}}{
	\abst{ 
		A single Majorana Kramers pair hosts only one component of the magnetic multipole. This can be used to determine the bulk Cooper-pair symmetry through surface-sensitive spectroscopic measurements, either by applying a magnetic field or by using a ferromagnet/superconductor junction.  
		This paper proposes that the electric response, which is free from the Meissner effect, can be used an alternative method to measure the bulk Cooper-pair symmetry in time-reversal-invariant superconductors with double Majorana Kramers pairs.
		The relationships among electric multipoles, strain tensors and superconducting symmetries under a given wallpaper group on the surfaces of topological crystalline superconductors are shown.  
		This study also reveals that only a specific irreducible representation of a uniform strain yields a gap in the double Majorana Kramers pairs for the topological-crystalline-superconductor candidate Sr$_3$SnO. This highlights the viability of electric detection regarding Cooper-pair symmetry.  
	}
 }{}

\begin{document}

\ifthenelse{\boolean{letter}}{}{\title{Electric Multipoles of Double Majorana Kramers Pairs}}

\ifthenelse{\boolean{letter}}{}{\author{Yuki Yamazaki}
	\affiliation{Department of Physics, Nagoya University, Nagoya 464-8602, Japan}
	\author{Shingo Kobayashi}
	\affiliation{RIKEN Center for Emergent Matter Science, Wako, Saitama 351-0198, Japan}
	\author{Ai Yamakage}
	\affiliation{Department of Physics, Nagoya University, Nagoya 464-8602, Japan}}

\maketitle

\ifthenelse{\boolean{letter}}{}{\section{Introduction}}
Majorana fermions are charge-neutral particles with identical particles and antiparticles \cite{Majorana}. They exist on the surfaces of topological superconductors (TSCs) as gapless Andreev bound states, and are completely stable as long as the superconducting gap remains \cite{Hu1526, Kashiwaya1641, Hasan3045, Qi1057, Tanaka011013, Sato076501, Haim2019a}. Majorana fermions are also charge-neutral on the surfaces of TSCs, and they obey non-Abelian statistics. These novel properties mean that Majorana fermions are expected to be applicable to fault-tolerant topological quantum computation \cite{Nayak1083}.

The topological nature of a three-dimensional TSC is characterised by its symmetry. Time-reversal symmetry (TRS) protects zero-energy gapless states on the surfaces of TSCs that form Kramers pairs; these are called Majorana Kramers pairs (MKPs). 
Superconducting doped topological insulators \cite{Hor057001, Fu097001, Sasaki217001, Sasaki217004, Hashimoto174527, Fu100509, Matano852, Yonezawa123}, and Dirac semimetals \cite{Aggarwal3237, Wang3842, Kobayashi187001, Hashimoto014510, Oudah13617, Kawakami041026} are promising candidates for time-reversal-invariant TSCs. On the other hand, crystalline symmetry combined with TRS defines a new type of TSC, called a topological crystalline superconductor (TCSC) \cite{ueno13, Taylor047006, zhang-kane-mele, chiu-yao-ryu, morimoto-furusaki, shiozaki14, Shiozaki195413, Benalcazar224503, Fang01944}. 

Majorana fermions that are topologically protected by crystalline symmetry have been shown to exhibit rich magnetic responses \cite{Sato094504, Chung235301, Nagato123603, shindou10, Mizushima12, Tsutsumi113707, Mizushima022001, shiozaki14}. 
For example, a crystalline-symmetry-protected MKP has been shown to exhibit a magnetic dipole (Ising) or octupole response, that is, a response distinct from conventional (complex) spin-$1/2$ fermions \cite{xiong17, kobayashi, yamazaki, kobayashi20}. 
Interestingly, the irreducible representation (irrep) of the magnetic multipole is the same as that of the bulk pair potential, which suggests that this could be a new direction for measuring the superconducting symmetry of TCSCs hosting a single MKP. 
However, to date only the magnetic response has been studied, and this may be difficult to use practically because of the Meissner effect. 
The electric response, however, could be more efficient regarding the detection of MKPs and the bulk superconducting symmetry. 

%For TCSCs with multiple MKPs, however, the electromagnetic response is only partially clear.
%Multiple MKPs can host magnetic multipoles, but they do not uniquely correspond to the bulk superconducting symmetry \cite{kobayashi}. 
%Additional measurements are therefore needed to determine the bulk pair potentials of TCSCs with multiple MKPs.
%The electric response is a potential measurement for resolving this issue[A5]. 
It has previously been revealed that double MKPs protected by crystalline symmetry can respond to an electric perturbation \cite{yamazaki07939}.
The electric responses of MKPs have also been classified for each superconducting pair potential when the symmetry of the surface is $pgg$ or $p4g$ wallpaper group (WG) \cite{kobayashi20}. 
Thus, although several properties of the electrical response of double MKPs have been clarified discursively, this study comprehensively clarifies new aspects of the electrical response from a different perspective.

This paper proposes that the electric response, which is unaffected by the Meissner effect and can be prepared more easily than a magnetic field, can determine the bulk Cooper-pair symmetry in TCSCs with double MKPs. 
The electric multipoles of double MKPs are related to the irreps of the bulk pair potential for a given surface WG. 
This study demonstrates that double MKPs on the surfaces of three-dimensional TCSC antiperovskites are gapped by an external uniform strain. 
It is useful to be able to identify bulk Cooper pairs electrically.  
Moreover, this result suggests that double MKPs on the surfaces of TCSCs can be controlled electrically. This would be well suited to nanofabrication and could potentially lead to the development of a novel Majorana device.

\ifthenelse{\boolean{letter}}{}{\section{Majorana Kramers pairs}}
\textbf{Internal, crystalline, and Cooper-pair symmetries.}
MKPs are zero modes localized on the surfaces of TSCs. 
Time-reversal and particle-hole symmetries stabilize a single MKP. 
On the one hand, multiple MKPs appear for a high-symmetry momentum when protected by the WG symmetry of the surface, in addition to time-reversal and particle-hole symmetries. These phenomena have been systematically classified \cite{chiu-yao-ryu, morimoto-furusaki, shiozaki14, Shiozaki195413, Fang01944, cornfeld19}. 
According to these classifications, double MKPs can only exist in magnetic chiral- and/or glide-plane-symmetric systems \cite{Shiozaki195413, Daido227001, yamazaki07939, yamazaki}, as described below.

Consider a three-dimensional Bogoliubov-de Gennes (BdG) Hamiltonian for a given momentum $\boldsymbol{k} = (k_x, k_y, k_z)$
\begin{align}
 H_{\mathrm{bulk}}(\boldsymbol{k})
 = \pmqty{
   h(\boldsymbol{k})-\mu & \Delta(\boldsymbol{k})
   \\
   \Delta(\boldsymbol{k})^\dag & -h(-\boldsymbol{k})^{\mathrm{T}} + \mu
 },
\end{align}
where $h(\boldsymbol{k})$ and $\Delta(\boldsymbol{k})$ are the Hamiltonian for the normal state and pair potential, respectively. 
$\mu$ denotes the chemical potential. 
The BdG Hamiltonian respects the time-reversal [$T H_{\mathrm{bulk}}(\boldsymbol{k}) T^{-1} = H_{\mathrm{bulk}}(-\boldsymbol{k})$, $T=-i s_y \mathcal K$] and particle-hole [$C H_{\mathrm{bulk}}(\boldsymbol{k}) C^{-1} = -H_{\mathrm{bulk}}(\boldsymbol{k})$, $C = \tau_x \mathcal K$] symmetries.
Here, $s_i$, $\tau_i$, and $\mathcal K$ are the $i$th Pauli matrices acting in the spin and Nambu spaces and the complex conjugation, respectively.
Chiral symmetry, $\Gamma = i C T$, which comprises the combined symmetry of time-reversal and particle-hole symmetries, is also preserved, $\{\Gamma, H_{\mathrm{bulk}}(\boldsymbol{k})\}=0$.
Crystalline systems have space-group (SG) symmetry.
We focus on the $(xy)$ surface and momentum located on $\boldsymbol{k} = (\boldsymbol{k}_0, k_z)$, where $\boldsymbol{k}_0 = (k_{0x}, k_{0y})$ is the time-reversal-invariant momentum.
The surface has a WG symmetry that is compatible with SG symmetry, $\mathrm{WG} \subset \mathrm{SG}$.  
The normal part of the Hamiltonian on a high-symmetry point $\boldsymbol{k}_0$ is invariant for $g$ a symmetry operation of WG, $[D_{\boldsymbol{k}_0}(g), h(\boldsymbol{k}_0, k_z)]=0$, where $D_{\boldsymbol{k}_0}(g)$ is the representation matrix of the little group.  
The pair potential is an order parameter and hence belongs (approximately) to an irrep of SG on (near) the phase transition point. 
For a one-dimensional irrep, the pair potential is either even parity ($\eta_g = 1$) or odd parity ($\eta_g=-1$) for $g$ as $D_{\boldsymbol{k}_0}(g) \Delta(\boldsymbol{k}_0, k_\perp) D_{\boldsymbol{k}_0}(g)^{\mathrm{T}} = \eta_g \Delta(\boldsymbol{k}_0, k_z)$.
The representation matrix $D_{\boldsymbol{k}_0}(g)$ of $g$ for the normal part is extended to $\tilde{D}_{\boldsymbol{k}_0}(g)$ in the Nambu space, as $\tilde{D}_{\boldsymbol{k}_0}(g) = D_{\boldsymbol{k}_0}(g) \oplus \eta_g D_{\boldsymbol{k}_0}^{*}(g)$, which commutes with the BdG Hamiltonian, $[\tilde{D}_{\boldsymbol{k}_0}(g), H_{\mathrm{bulk}}(\boldsymbol{k}_0, k_z) ]=0$ for both even and odd parity pairings.  
The particle-hole and chiral transforms also commute with $\tilde D(g)$, $[C,  \tilde{D}_{\boldsymbol{k}_0}(g)] = [\Gamma, \tilde{D}_{\boldsymbol{k}_0}(g)] =0 $ for even-parity pairings $\eta_g =1$. 
This study explicitly considers only one-dimensional irreps, because higher-dimensional irreps are regarded as one-dimensional irreps of the subgroup. 

As $[\tilde D_{\boldsymbol{k}_0}(g), H_{\mathrm{bulk}}(\boldsymbol{k}_0, k_z)]=0$, 
they are block-diagonalized as $U(g)^\dag H_{\mathrm{bulk}}(\boldsymbol{k}_0, k_z) U(g) = H_{\mathrm{bulk}}^1(\boldsymbol{k}_0, k_z) \oplus \cdots \oplus H_{\mathrm{bulk}}^{|g|}(\boldsymbol{k}_0, k_z)$ and $U(g)^\dag \tilde D_{\boldsymbol{k}_0}(g) U(g) = \omega_1(g) 1_{\dim{H_{\mathrm{bulk}}^1(\boldsymbol{k}_0, k_z)}} \oplus \cdots \oplus \omega_{|g|}(g) 1_{\dim{H_{\mathrm{bulk}}^{|g|}(\boldsymbol{k}_0, k_z)}}$, where $|g|$ denotes the order of $g$, $1_{n}$ denotes the $n \times n$ identity matrix, and the $j$th eigenvalue of $\tilde{D}_{\boldsymbol{k}_0}(g)$ is given by 
\begin{align}
 \omega_j(g) = e^{-i \pi (2j-1)/|g|}, \quad j=1, \cdots, |g|,
\end{align}
for rotations and mirror reflections. 
The glide plane $g=\{m_g | \boldsymbol{\tau}_g\}$, where the Seitz notation is adapted to apply to symmetry operations, that is, the mirror $m_g$ followed by the translation $\boldsymbol{\tau}_g$, on the Brillouin zone boundary with $\boldsymbol{k}_0 \cdot 2 \boldsymbol{\tau}_g = \pi$ is squared to be $\tilde{D}_{\boldsymbol{k}_0}(g)^2 = - e^{-i \boldsymbol{k}_0 \cdot 2 \boldsymbol{\tau}_g} = 1$ and has the eigenvalues $\omega_1(g)=1$ and $\omega_2(g)=-1$.  
%
%When $g'$ is not commuting with $g$, the same procedure is conducted on a different basis. 
%

\textbf{Topological invariants for double MKPs.}
For symmorphic symmetry operations, the time reversal of $H^i_{\mathrm{bulk}}(\boldsymbol{k}_0, k_z)$ is  $H^{|g|-i+1}_{\mathrm{bulk}}(\boldsymbol{k}_0, -k_z)$ owing to $\omega_i(g)^* = \omega_{|g|-i+1}(g)$.
When the pair potential is even-parity for $g$, $\eta_g = 1$, $H^i_{\mathrm{bulk}}(\boldsymbol{k}_0, k_z)$ maintains the chiral symmetry. 
Therefore, $H^i_{\mathrm{bulk}}(\boldsymbol{k}_0, k_z)$ is in class AIII and is characterised by the winding number $w^i[g] \in \mathbb Z$, which corresponds to the number of Majorana fermions generated from $H^i_{\mathrm{bulk}}(\boldsymbol{k}_0, k_z)$. 
For the glide ($g$) plane, the decomposed Hamiltonian still preserves time-reversal symmetry, that is, $H^i_{\mathrm{bulk}}(\boldsymbol{k}_0, k_z)$ belongs to class DIII and is characterised by the $\mathbb Z_2$ invariant $\nu^i[g]$ because $\omega_1(g)=1$ and $\omega_2(g)=-1$ are real.  

We define the number of MKPs protected by the chiral symmetry as 
\begin{align}
 N[g] = \frac{1}{2}\sum_{i=1}^{|g|} \qty|w^i[g]|,
\end{align}
for $g$ a symmorphic symmetry operation.
$N[g]$ can be verified as an integer because $|w^i[g]| = |w^{|g|-i+1}[g]|$ holds due to TRS. 
In contrast, for nonsymmorphic glide-plane-($g$)-symmetric systems, single or double MKPs can appear on the surface for $(\nu^1[g], \nu^2[g])=(1,0),(0,1)$, or $(\nu^1[g], \nu^2[g]) = (1,1)$, respectively.  
As a result, double MKPs can exist in symmorphic systems when $N[g]=2$ and glide-plane-symmetric systems for $\nu^1[g]=\nu^2[g]=1$. 
For example, the former occurs in systems with double Fermi surfaces, and the latter occurs in $\nu^1[g]=\nu^2[g]$, when enforced by additional symmetries.

The topological invariants defined above are not only determined by the properties of $g$, but are also restricted by $g'$ ($\ne g$). 
Table \ref{double} shows the results of considering all of these conditions for the WGs. This indicates which irreps of bulk pair potentials induce double MKPs protected by $w^i[g]$ and/or $\nu^i[g]$ \cite{kobayashi}. 
Double MKPs are protected by $N[g]=2$ for symmorphic WGs, except for threefold rotation and $\nu^1[g]=\nu^2[g]=1$ for some cases in nonsymmorphic WGs. 
$p3$ with an $A$ pairing can create double MKPs for two cases: $N[g]=2$ or $N[g]=\nu^2[g]=1$. The latter case is understood as follows.  
When $g$ is the threefold rotation ($C_3$), $H^i_{\mathrm{bulk}}(\boldsymbol{k}_0, k_z)$ for $i=1$ and $3$ belong to AIII, whereas $H^2_{\mathrm{bulk}}(\boldsymbol{k}_0, k_z)$ is in class DIII because $\omega_2(C_3)^* = \omega_2(C_3)$. 
The former ($H^1$ and $H^3$) is characterised by the winding number $w^i[C_3]$ ($i=1, 3$), whereas the latter ($H^2$) is characterised by the $\mathbb Z_2$ invariant $\nu^2[C_3]$.
The same holds for $p31m (A_2)$, $p3m1 (A_2)$, $p6 (A)$, and $p6m (A_2)$.

\begin{table}
	\caption{Topological invariants (Topo) of double MKPs and irrep of bulk pair potential projected onto the surface for WGs.
	The momenta are located on the high-symmetry points $B$, $X$, $S$, and $M$, as defined in the Bilbao Crystallographic Sever \cite{aroyo14, tasci12}, for the nonsymmorphic WGs $pg$, $pmg$, $pgg$, and $p4g$, respectively.
 $pgg$ and $p4g$ on the high-symmetry point can host either zero or double MKPs; they cannot host a single MKP. }
\ifthenelse{\boolean{letter}}{}{\begin{ruledtabular}} 
 	\begin{tabular}{ll}
 		Topo & WG (irrep)
 		\\
 		\hline
 		$N[g]=2$ & $p2$ ($A$), $p3$ ($A$), $p4$ ($A$), $p6$ ($A$), 
 		$pm$ ($A'$),  
 		\\ &
 		$pmm$ ($A_2$, $B_1$, $B_2$), 
 		$p31m$ ($A_1$, $A_2$), 
 		\\ &
 		$p3m1$ ($A_1$, $A_2$), 
% 		\\ &
 		$p4m$ ($A_2$), 
 		\\ &
     	$p6m$ ($A_2$, $B_1$, $B_2$), 
     	\\ & $pmg$ ($A_1$), $pgg$ ($A_2$), 
% 		\\ &
 		$p4g$ ($A_2$)
 		\\
 		$\nu^1[g]=\nu^2[g]=1$ & $pg$ ($A'$), $pmg$ ($B_1$), $pgg$ ($A_1$, $B_1$, $B_2$), 
 		\\ &
 		$p4g$ ($A_1$, $B_1$)
 		\\
 		$N[C_3]=\nu^2[C_3]=1$ & $p3 (A)$, $p31m (A_2)$, $p3m1 (A_2)$, 
 		\\ &
 		$p6 (A)$, $p6m (A_2)$
 	\end{tabular}
\ifthenelse{\boolean{letter}}{}{\end{ruledtabular}} 
\label{double}
%\vspace{-3em}
\end{table}

\textbf{Electric degrees of freedom of double MKPs.}
Hereafter, the relationships between the electric response, crystalline symmetry, and Cooper pair symmetry are explained in terms of the effective surface theory for MKPs with time-reversal, particle-hole, and WG symmetries.
An effective theory for $N$ MKPs, zero modes, around the time-reversal-invariant momentum on the surface $\boldsymbol{x}=(x,y)$ can be constructed using Majorana field operators $\psi_{s}(\boldsymbol{x})$, $s=1, \cdots, 2N$. This satisfies the self-conjugate condition $\psi_{s}^\dag(\boldsymbol{x}) = \psi_{s}(\boldsymbol{x})$ on an appropriate basis.  
Assuming that $\psi_{2s-1}(\boldsymbol{x})$ and $\psi_{2s}(\boldsymbol{x})$ form a Kramers pair without loss of generality, time-reversal is represented by $T_{\mathrm{surf}} = (-i s_y) \oplus \cdots \oplus (-is_y)$.
The field operators obey the anticommutation relationship $\{ \psi_{s}(\boldsymbol{x}), \psi_{s'}(\boldsymbol{x}') \} = \delta_{ss'}\delta^2(\boldsymbol{x} - \boldsymbol{x}')$. 
The Hamiltonian on the surface effective theory is induced by a uniform external field $F$ as follows: 
\begin{align}
 H_{\mathrm{surf}, \mathrm{ex}} = - O F,
 \
 O =
 \frac{1}{2} \int d^2x \sum_{ss'} \psi_{s}(\boldsymbol{x}) (A_F)_{ss'} \psi_{s'}(\boldsymbol{x}),
 \label{OF}
\end{align}
where $A_F$ is conjugate to $F$ and should be given by an antisymmetric Hermite matrix. 
For a single MKP, $N=1$, only one antisymmetric Hermite matrix is proportional to $s_y$.
This means that a single MKP hosts only one magnetic (time-reversal-odd) operator. 
This is in sharp contrast to conventional (complex) fermions, $c_s(\boldsymbol{x})$, which always have three-component magnetic dipoles $\sum_{ss'} c_s^\dag(\boldsymbol{x}) (s_i)_{ss'} c_{s'}(\boldsymbol{x})$ for $i=x$, $y$, and $z$. 
On the other hand, double MKPs, for which $N=2$, have six operators that are represented by six antisymmetric Hermite matrices: $s_y \sigma_0$, $s_y \sigma_x$, $s_y \sigma_z$, $s_0 \sigma_y$, $s_x \sigma_y$, and $s_z \sigma_y$.  
The former four are time-reversal-odd (magnetic), $T_{\mathrm{surf}} A_F T_{\mathrm{surf}}^{-1} = -A_F$, and the latter two are time-reversal-even (electric), $T_{\mathrm{surf}} A_F T^{-1}_{\mathrm{surf}} = A_F$. 

Owing to symmetry constraints, $O$ and $F$ must belong to the same irrep of WG. 
This condition enables the determination of the symmetry of the pair potential from their response to an external field, as follows: 
Table \ref{double} shows the relationship between the topological invariants and the irrep of the pair potential: if a topological invariant, $N[g]$ and/or $\nu^i[g]$, is nonzero, then the pair potential is even parity for $g$. 
A static uniform perturbation that is an odd parity for $g$ makes $N[g]$ and $\nu^i[g]$ ill defined. This creates a gap in the double MKPs. 
This leads that double MKPs with $N[g]$ and/or $\nu^i[g]$ hosts electric multipole operators $O$ being odd parity for $g$. 
Consequently, the symmetry of the pair potential is related to the electric multipoles (and the electric response to an external field, Eq.~(\ref{OF})) of the double MKPs. 

Table \ref{electric} shows a representative example for $p4m$ with $A_2$ and $pg$ with $A'$ pairings, where the double MKPs are protected by $N[g]=2$ and $\nu^1[g]=\nu^2[g]=1$, respectively (Table \ref{double}). 
\begin{table}
	\caption{Electric multipoles of double MKPs. 
		They emerge on the surface with WG symmetry when the bulk pair potential belongs to the irrep ($\Delta$) of the point group corresponding to WG. 
		They have topological invariants (Topo) $N[g]=2$ protected by the magnetic chiral symmetry of $g$ or $\nu^1[g]=\nu^2[g]=1$ protected by glide-plane symmetry.  
		The resulting electric operators of double MKPs are decomposed into irreps (Electric). 
		The last column shows the strain tensor belonging to the same irrep of Electric as a representative electric perturbation.
		The irreps of $C_{4v}$ and $C_s$, which are compatible with $p4m$ and $pg$, respectively, are defined in Table \ref{character}.
		In $pg$, the glide plane is set to the $(xz)$ plane. 
	}
		\begin{tabular}{lllll}
			WG & $\Delta$ & Topo & Electric & Strain
			\\
			\hline
			$p4m$ & $A_2$ & $N[C_4]=N[C_2]=2$ & $E$ & $(u_{xz}, u_{yz})$,
			\\ & & & & $(u_{zx}, u_{zy})$
			\\
			$p4m$ & $A_2$ & $N[C_4]=2$ & $B_1+B_2$ & $u_{xx}-u_{yy}$,
			\\ & & & & $u_{xy}+u_{yx}$
			\\
			$pg$ & $A'$ & $\nu^1[g]=\nu^2[g]=1$ & $2A''$ & $u_{xy}$, $u_{yx}$,
			\\ & & & &  $u_{yz}$, $u_{zy}$
		\end{tabular}
	\label{electric}
\end{table}
\begin{table}
	\caption{Character table for $C_{4v}$ and $C_s$ compatible with $p4m$ and $pg$, respectively. }
		\begin{tabular}{llllll|lll}
			$C_{4v}$ & $E$ & $2C_4(z)$ & $C_2$ & $2\sigma_v$ & $2\sigma_d$ & $C_{s}$ & $E$ & $\sigma_h(xz)$
			\\
			\hline
			$A_1$ & 1 & 1 & 1 & 1 & 1 & $A'$ & 1 & 1
			\\
			$A_2$ & 1 & 1 & 1 & $-1$ & $-1$ & $A''$ & 1 & $-1$
			\\
			$B_1$ & 1 & $-1$ & 1 & 1 & $-1$ &
			\\
			$B_2$ & 1 & $-1$ & 1 & $-1$ & 1 &
			\\
			$E$ & 2 & 0 & $-2$ & 0 & 0 &
		\end{tabular}
\label{character}
\end{table}
The first case in Table \ref{electric} is for double MKPs protected by the magnetic chiral symmetry of $C_2$ and $C_4$.
They have the electric operators breaking $C_2$ and $C_4$ symmetries, that is, the $E$ irrep defined in Table \ref{character}. 
$A_1$, $A_2$, $B_1$, and $B_2$ electric irreps respect $C_2$ symmetry and are not coupled to the double MKPs. 
The second case in Table \ref{electric} is protected solely by the magnetic chiral symmetry of $C_4$. 
$C_4$-symmetry-breaking electric operators are coupled to the double MKPs, that is, $B_1$ and $B_2$ irreps (Table \ref{character}), which exhibit odd parity for $C_4$ and even parity for $C_2$. 
The double MKPs for $pg$ are protected by the glide-plane symmetry.
$A''$ electric irrep, which exhibits odd parity for the glide plane (Table \ref{character}), is coupled to the double MKPs. 
This relationship can be clarified for all of the WGs using the general theory of Majorana multipoles \cite{kobayashi20}, which is shown in \ifthenelse{\boolean{letter}}{Supplemental Material \cite{suppl}}{Appendix \ref{id}}.

Strain is a representative electric perturbation that is coupled to electric quadrupoles.  
The unsymmetrized strain tensor $u_{ij} = \partial_i u_j$, where $u_i$ is a vector field called the lattice displacement field; it is transformed by $\boldsymbol{x} \to g \boldsymbol{x}$ as $\boldsymbol u \to g \boldsymbol u$, which is decomposed into irreps $\Gamma_i$ of a WG, $u_{\Gamma_i}$. 
Electric multipoles that is conjugate to the strain also decompose into the irreps $O_{\Gamma_1}$ and $O_{\Gamma_2}$; thus, $H_{\mathrm{surf}, \mathrm{ex}} = - \sum_{i=1}^2 O_{\Gamma_i} u_{\Gamma_i}$. 

\textbf{Application to Sr$_3$SnO.}
The proposed theory is applied to an antiperovskite.
Antiperovskite $A_3BX$, $A=\mathrm{Ca}, \mathrm{Sr}, \mathrm{La}$, $B=\mathrm{Pb}, \mathrm{Sn}$, $X=\mathrm{C}, \mathrm{N}, \mathrm{O}$, is a candidate material for zero-gap semiconductors for small spin-orbit coupling, or for topological crystalline insulators, in un-doped cases \cite{kariyado11, kariyado12, hsieh14}.
Interestingly, Sr$_3$SnO, becomes a superconductor at temperatures below 5~K \cite{Oudah13617}, and could possibly be an unconventional superconductor with strongly hybridised orbitals \cite{Kawakami041026}. 
Here it is shown that a uniform strain induces a gap in double MKPs for a possible TCSC state of Sr$_3$SnO. 
%%%%%%%%%%%%%%%%%%%%%%%%%%

Consider a model for an antiperovskite with a pair potential of $A_{1u}$ of the $O_h$ point group.  
The bulk Hamiltonian for $\boldsymbol{k}=(k_x, k_y, k_z)$ is given by \cite{Kawakami041026}
$H_{\mathrm{bulk}}(\boldsymbol{k}) = h(\boldsymbol{k})\tau_z + \Delta_0 \sigma_x \tau_x$ with
\begin{align}
h(\boldsymbol{k}) &= \qty[-m_0 + \alpha \sum_{i=x,y,z} \{2- 2\cos(k_i)\} ] \sigma_z, 
 \notag\\ & \quad 
 + \sin \boldsymbol{k} \cdot \qty(v_1 \boldsymbol{J} + v_2 \tilde{\boldsymbol{J}})\sigma_x - \mu \sigma_0,
\end{align}
where $\sigma_0$ and $\sigma_i$ are the identity and the $i$th Pauli matrix for the orbital degrees of freedom, respectively, and $\boldsymbol{J}$ and $\tilde{\boldsymbol{J}}$ are $4 \times 4$ the matrices of spin $J=3/2$. These are related to $\tilde{J}_i \equiv \frac{5}{3}\sum_{j \neq i}J_j J_i J_j - \frac{7}{6}J_i$. 
The explicit representations of $\boldsymbol{J}$ and $\tilde{\boldsymbol{J}}$ and the symmetry operations in this system are shown in \ifthenelse{\boolean{letter}}{Supplemental Material \cite{suppl}}{Appendix \ref{angular}}.

The proposed general theory can identify possible couplings between a perturbation and double MKPs on the (001) surface that respect the $p4m$ ($C_{4v}$) symmetry.
For $-1/3 < v_2/v_1 < 1/2$, the winding numbers are given by $w^{1}[C_4]=-1$ and $w^4[C_4]=1$ in the $D(C_4)=e^{\mp i \pi/4}$ eigenspaces, and by $w^{2}[C_4]=-1$ and $w^3[C_4]=1$ in the $D(C_4)=e^{-\mp i 3\pi/4}$ eigenspaces \cite{Kawakami041026}. These are contributed by the $J=1/2$ and $J=3/2$ Fermi surfaces, respectively. 
This results in $N[C_4]=2$.
In contrast, the winding number of $C_2$ vanishes, $w^{1}[C_2] = w^{2}[C_4] + w^{4}[C_4]= 0$ and $w^2[C_2] = w^{1}[C_4] + w^{3}[C_4] = 0$.  
As a result, this system can be classified into the second case in Table \ref{electric}. 
These double MKPs have electric quadrupoles of $B_1+B_2$ of $C_{4v}$ on the (001) surface, which are coupled to $u_{xx}-u_{yy}$ and $u_{xy}+u_{yx}$.

To verify this coupling, the diagonal strains $u_{xx}$, $u_{yy}$, and $u_{zz}$ are applied. These strains decompose into $A_{ 1 g }$ [$u_{xx}+u_{yy}+u_{zz}$] and $E_g$ [$(2u_{zz}-u_{xx}-u_{yy}, u_{xx}-u_{yy})$] for irreps of $O_h$. 
The former renormalizes the parameters, such as the Fermi level. 
The latter is coupled to the operators (matrix) belonging to the $E_g$ irrep as
\begin{align}
	H_{\mathrm{bulk}, E_g} &= 
\alpha_{\mathrm E} \qty[\qty(2J_z^2 -J_x^2 -J_y^2) \qty(2u_{zz}-u_{xx}-u_{yy}) ]\tau_z \notag\\&\quad
+ \alpha_{\mathrm E} \qty[3 \qty(J_x^2 -J_y^2)\qty(u_{xx}-u_{yy}) ]\tau_z,
\end{align}
which is invariant for the symmetry operation $g$ of $O_h$ as $D(g) H_{\mathrm{bulk}, E_g} D(g)^\dag = H_{ \mathrm{bulk}, E_g}|_{u_{ij} \to g_{ii'} g_{jj'} u_{i'j'}}$.
The first and second terms contain the strains of $A_1$ ($u_{zz}$ and $u_{xx}+u_{yy}$) and $B_1$ ($u_{xx}-u_{yy}$) irreps of $C_{4v}$ on the (001) surface. 
The proposed general theory predicts that the latter term induces a gap in the double MKPs on the surface. 
%These two terms are denoted here as $H_{A_1}$ and $H_{B_1}$. 

\begin{figure}
	\centering
	\includegraphics[scale=0.28]{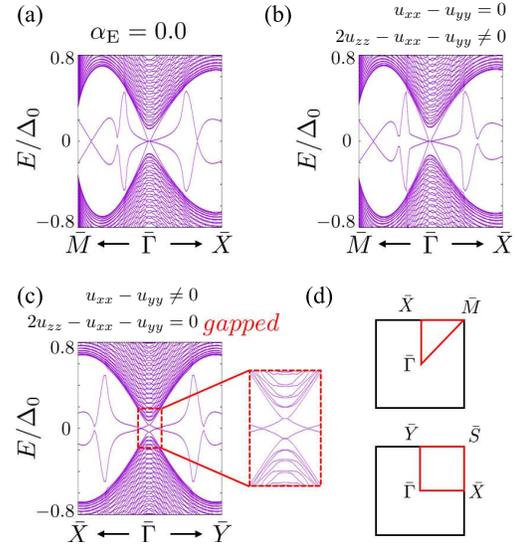}
	\caption{Energy spectrum of the finite-sized BdG Hamiltonian for Sr$_3$SnO. (a) Without a strain, double MKPs appear at the $\bar\Gamma$ point. 
		They are still gapless under the strain with (b) $u_{xx}-u_{yy}=0$ while they are gapped with $2u_{zz}-u_{xx}-u_{yy}=0$. 
		The energy spectra are drawn along the high-symmetry lines indicated in (d). 
%		These results are consistent with those shown in Table \ref{electric}.
 The parameters are as follows: $m_0=1.0$, $\mu=-1.25$, $\alpha=0.64$, $v_1=1.0$, $v_2=-0.2$, $\Delta_0=0.025$, $\alpha_{\mathrm{E}}=0.02$. $2u_{zz}-u_{xx}-u_{yy}=1$ for (b), and $u_{xx}-u_{yy}=1$ for (c). 
}
	\label{p4m}
%	\vspace{-2em}
\end{figure}

The energy spectrum of the finite-sized model for Sr$_3$SnO is calculated with the $(001)$ surface. 
%%%%%%%%%%%%%%%%%%%%%%%%%%%%%%%%
In the absence of the strain, double MKPs emerge at the $\bar\Gamma$ point ($k_x=k_y=0$), as shown in Fig.~\ref{p4m}(a); they are protected solely by fourfold rotational symmetry. 
The diagonal strain $2u_{zz} -u_{xx} -u_{yy}$ under the condition $u_{xx}-u_{yy}=0$, which belongs to the $A_1$ irrep of $C_{4v}$, does not affect the double MKPs because it maintains the symmetry of the system [Fig.~\ref{p4m}(b)]. 
They are gapped by the strain of $u_{xx} - u_{yy} \ne 0$, which belongs to the $B_1$ irrep, as shown in Fig.~\ref{p4m}(c). 
This is consistent with the general results summarized in Table \ref{electric}. 
%This also confirms that the general theory accurately describes the strain response of UCoGe by hosting a time-reversal-invariant topological pairing \cite{suppl}.  

\ifthenelse{\boolean{letter}}{}{\subsection{UCoGe (\texorpdfstring{$pg$}{pg})}}

\ifthenelse{\boolean{letter}}{\textbf{Discussion.}}{\section{Summary}}
Here, electric multipoles of double MKPs and their coupling to strains on the surface of TCSCs were established.
The relations among bulk superconducting symmetries and strain tensors coupled to double MKPs have been systematically shown. 
This can easily predict the effects of applying strain on the surfaces of TCSCs. 
The most straightforward application is a spatially uniform static strain, which can create a gap in the double MKPs hosting the corresponding electric multipoles. Here, this is demonstrated for a model of Sr$_3$SnO. 
These results suggest that spectroscopic measurements taken under strain can detect double MKPs on the surface. 
In addition, the electric response may be more experimentally accessible because, unlike the magnetic response, it is presumed to be free from the Meissner effect.
Therefore, complementary magnetic and electric measurements should be used to determine the symmetry of the bulk pair potential in TCSCs with double MKPs.

This study only considered the zero-order coupling between double MKPs and strain, and it was assumed that the coupling constant was finite.
This satisfied the symmetry requirement, and such a coupling inevitably exists. Furthermore, its effective model was able to describe various phenomena. However, the scope of the effective model does not permit quantitative evaluation of the microscopic mechanism of the coupling between the strain and MKP, or the coupling constant. This evaluation would require a return to higher-energy models, that is, bulk theory. Such a study will be essential for future progress in MKP research. This is an important issue that should be addressed in future studies.

The results presented here also imply a possible coupling between double MKPs and non-uniform dynamic strains, that is, ultrasounds, $
	H_{\text{sound}}(t) =  -\int d^2x \sum_{i = 1}^2 
	\alpha_{\Gamma_i}
	 \psi(\boldsymbol{x}, t)^{\mathrm{T}} 
	 A_{\Gamma_i} 
	 \psi(\boldsymbol{x}, t) u_{\Gamma_i}(\boldsymbol{x}, t)
$ with $\alpha_{\Gamma_i}$ as the coupling constant, which may induce exotic charge/heat and spin transport carried by MKPs. Strains are expected to drive spin currents because they share the same symmetry (irrep). 
This shall be addressed in a future study. 

\begin{acknowledgments}
	The authors are grateful to Masatoshi Sato for fruitful discussions. 
	A.Y. was supported by JSPS KAKENHI Grant Nos.~JP20K03835 and JP20H04635. 
	S.K. was supported by the JSPS KAKENHI Grant No.~JP19K14612, and by the CREST project (JPMJCR16F2,~JPMJCR19T2) from the Japan Science and Technology Agency (JST).
\end{acknowledgments}

\bibliography{ref}
\clearpage
\end{document}